\documentclass[prl,twocolumn,aps, preprintnumbers,showpacs, showkeys,amsmath]{revtex4}
\usepackage{color}
\usepackage[active]{srcltx}
\usepackage{amsmath,amsfonts,amssymb,amsthm,amstext,amscd,eucal,srcltx}
\usepackage{epsfig,graphicx,bm}
\usepackage{epstopdf, epsf}
\usepackage{dcolumn}
\usepackage{hyperref}

\newcommand{\mpl}{M_{\rm pl}}
\def\tchi{\tilde\chi}
\newcommand{\be}{\begin{equation}}
\newcommand{\ee}{\end{equation}}

\newcommand{\bse}{\begin{subequations}}
\newcommand{\ese}{\end{subequations}}
\newcommand{\bea}{\begin{eqnarray}}
\newcommand{\eea}{\end{eqnarray}}
\newcommand{\ba}{\begin{array}}
\newcommand{\ea}{\end{array}}
\newcommand{\bc}{\begin{center}}
\newcommand{\ec}{\end{center}}

\begin{document}
\preprint{IPM/P-2012/009}  
\vspace*{3mm}
\title{Gauge-flation Vs Chromo-Natural Inflation }%
\author{M.~M.~Sheikh-Jabbari}
\affiliation {School of Physics, Institute for research in fundamental
sciences (IPM), P.O.Box 19395-5531, Tehran, Iran }

\begin{abstract}
Gauge-flation, non-Abelian gauge field inflation, which was introduced in \cite{gauge-flation-short} and analyzed more  thoroughly in \cite{gauge-flation-long}, is a model of inflation driven by non-Abelian gauge fields minimally coupled to Einstein gravity. In this model certain rotationally invariant combination of gauge fields play the role of inflaton. Recently, the chromo-natural inflation model was proposed \cite{chromo-natural} which besides the non-Abelian gauge fields also involve an axion field. In this short note we show that the model involving axions, indeed allows for various slow-roll trajectories for different values of its parameters: A specific trajectory discussed in \cite{chromo-natural} starts from a ``small axion'' region, while the trajectory considered in \cite{gauge-flation-short,gauge-flation-long} corresponds to a ``large axion'' region. 

\end{abstract}
\maketitle

Given the current cosmological  data \cite{CMB-data}, inflation \cite{Inflation-Books} has appeared as the leading paradigm for the early Universe cosmology. The data indicate an upper bound for the Hubble parameter $H$ during inflation $H\lesssim 10^{-5}\mpl\simeq 10^{13}-10^{14}$ GeV. For the vast majority  of inflationary models,  inflationary dynamics is driven by one or more scalar fields coupled to gravity and usually inflation takes place in a ``slow-roll'' regime such that during inflation we have an almost constant effective potential of order (or smaller than) $\mu\sim \sqrt{H\mpl}\lesssim 10^{16}$ GeV. During slow-roll the kinetic energy should remain small compared to the potential terms.

Motivated with the above bounds on $H$ it is  natural to seek building inflationary model within beyond-standard particle physics models. In these theories, although we have scalar fields,  e.g.  Higgs fields required for gauge symmetry breaking, the potential for the Higgs fields are not generically  flat enough to lead to slow-roll dynamics. (See, however, \cite{Higgs-Inflation} for ways to get round this issue.)

On the other hand, all particle physics models are gauge field theories and one may try to use gauge fields, rather than scalars, as tools for inflationary model building. This idea, to start with, has two basic obstacles, an observational and a theoretical one: 1) The Universe at large scales is homogenous and isotropic and turning on a vector gauge field will introduce a preferred direction in the Universe, breaking the isotropy and, 2)  to have a consistent (ghost free) theory of gauge fields we need gauge symmetry (which  may be consistently broken via Higgs mechanism). The energy momentum tensor of Yang-Mills theory, regardless of the gauge field configuration turned on, is traceless with $\rho=3P$ ($\rho\geq 0$ and $P$ are energy density and pressure associated with gauge field  configuration), while for having inflation we need $\rho+3P<0$.

As explained and reviewed in \cite{gauge-flation-short,gauge-flation-long} these two obstacles may be overcome in \emph{non-Ableian} gauge theories. Let us take any non-Abelian gauge group $G$. $G$ necessarily  has at least one $SU(2)$ subgroup (it may have more than one such $SU(2)$ subgroup). The global part of this $SU(2)$ subgroup may be identified with the space rotation group, which is again an $SU(2)$. Since gauge fields are defined up to gauge transformations we will have a way to restore rotation symmetry despite turning on a gauge field in the background.
(This idea in the context of cosmology,  though not for inflationary model building, has been pointed out and discussed in \cite{Galtsov}.)

The second obstacle, may be overcome either by considering gauge theories beyond simple Yang-Mills, or addition of other matter fields coupled to Yang-Mills. The first idea was examined and studied in \cite{gauge-flation-short,gauge-flation-long}, by addition of an $(F\wedge F)^2$ to the Yang-Mills action, $F$ is the gauge field strength two-from. The second idea was recently studied in \cite{chromo-natural} by adding an axion field $\chi$, coupled to the Yang-Mills theory via standard $\chi F\wedge F$ interaction term.

The purpose of this short note is elucidate the connection between the two models, especially focusing on the slow-roll inflationary trajectories of the two models.
More explicitly, we obtain the gauge-flation action of \cite{gauge-flation-short} once we integrate out the axion field $\chi$,  when axion is close to the minimum of its potential and show that the integrated out theory provides a very good description around the slow-roll trajectory path. As we will argue, the slow-roll trajectory discussed in \cite{chromo-natural} can occur for a wider range of initial values of the axion field, where axion is not necessarily close to the minimum of its potential. As shown in \cite{Mark-Peter},  for the inflationary path of \cite{chromo-natural} there is an upper bound on the initial value of axion field leading to successful inflation.

\textbf{\emph{Non-Abelian gauge fields in inflationary setups}}

Consider an $su(2)$ gauge field $A^a_{~\mu}$, where
$a,b=1,2,3$ and $\mu,\nu=0,1,2,3$ are respectively used for
the gauge algebra and space-time indices. The gauge field strength $F$ is
\be\label{F-general}%
F^a_{~\mu\nu}=\partial_\mu A^a_{~\nu}-\partial_\nu
A^a_{~\mu}-g\epsilon^a_{~bc}A^b_{~\mu}A^c_{~\nu}, %
\ee%
where $\epsilon_{abc}$ is the totally antisymmetric tensor, the structure constant of $su(2)$ algebra.
The chromo-natural inflation model of \cite{chromo-natural} is described by the action
\be\label{chromo-narutal-action}
\begin{split}
{\cal L}_{c.n.} =-\bigg(&\frac{R}{2}+\frac{1}{4}F^a_{~\mu\nu}F_a^{~\mu\nu}+\frac12(\partial_\mu\tchi)^2\cr & +\mu^4(1+\cos\frac{\tchi}{f})
+\frac{\lambda}{8f}\tchi F\tilde F\biggr)\,
\end{split}\end{equation}
where we have set $8\pi G\equiv \mpl^{-2}=1$, $\mu, f$ are parameters of dimension of energy, $\lambda$ is an order one dimensionless coupling and
\be
F\tilde F=\epsilon^{\mu\nu\alpha\beta} F_{\mu\nu}^a F_{\alpha\beta}^a\ .
\ee
As discussed in \cite{chromo-natural}, slow-roll inflationary dynamics for the above action can happen for $\mu, f \ll \mpl$, only if we also turn on a ``rotationally invariant gauge field'' of the form \cite{gauge-flation-short,gauge-flation-long,Galtsov}
\be\label{A-ansatz-background}
A^a_{~\mu}=\left\{
\begin{array}{ll} a(t)\psi(t)\delta^a_i\, ,\qquad  &\mu=i
\\
0\,, \qquad &\mu=0\,,
\end{array}\right.
\ee%
in the temporal gauge. Moreover, slow-roll dynamics requires both $\psi$ and $\tchi$ fields to be slowly rolling:
$\dot\tchi/H\tchi, \dot\psi/H\psi \ll 1$.

We next note that the potential for the axion field is of the form
\be\label{axion-potential}
U(\tchi,\Xi)=\mu^4\left(1+\cos\frac{\tchi}{f}+\Xi\frac{\tchi}{f}\right)\,,
\ee
where $\Xi= \frac{\lambda}{8\mu^4 } F\tilde F$.
This potential has an extremum at $\tchi=\chi$:
\be\label{axion-minimum}
\sin\frac{\chi}{f}={\Xi}\simeq \frac{3g\lambda}{\mu^4}\ H\psi^3\,.
\ee
Second equality in the above is written in the leading order in slow-roll parameter, dropping $\dot\psi$ against $H\psi$. The potential has extrema if $\Xi\leq 1$. This latter puts an upper bound on $\psi$ (for given $g$ and $H$).\footnote{It is worth noting that in the discussions of \cite{chromo-natural} \eqref{axion-minimum} has been viewed as the condition minimizing effective potential for the $\psi$ field.} Eq.\eqref{axion-minimum} has two solutions, $\chi/f$ and $\pi-\chi/f$. Here we choose our conventions such that $\chi/f\in (\pi/2,\pi)$, i.e. $\chi/f$ corresponds to minimum while $\pi-\chi/f$ to maximum.\\

\textbf{\emph{Gauge-flation from expanding around minimum of the axion potential---}}
As in standard quantum field theory treatments one may expand the axion theory around its minimum of the potential at $\tchi=\chi$ given in \eqref{axion-minimum} and integrate out the fluctuations \cite{Peskin}. This may be done writing $\tchi=\chi+\delta\chi$, expanding the action up to the second order in $\delta\chi$ and computing the one loop effective action/potential:
\be\label{one-loop-potential}
U_{eff}(\chi)=U(\chi)+ \frac{i}{2}\ln \det(\Box+\frac{\mu^4}{f^2}\cos\frac{\chi}{f})\,,
\ee
where $U(\chi)$ is the potential \eqref{axion-potential} computed at the minimum $\tchi=\chi$ and  $\Box$ is the Laplacian computed on the inflationary (almost-de Sitter) background.

To compute the $\ln\det$ term explicitly we note that during inflation $\mu^4\sim 3H^2\mpl^2$ and that $f^2\sim \mu\mpl$ \cite{chromo-natural} and hence $\mu^4/f^2\sim f^2\cdot H/\mpl\ll f^2$. The $\Box$ term may be approximated by its
flat space value once we recall that the main contribution comes from subhorizon momenta. The integral over the momenta  may be cut off at UV cutoff $\Lambda$ which may be taken equal to $f$. Using equations in \cite{Peskin}, and in the $\mathrm{\overline{MS}}$ scheme, we find
$$
U_{eff}(\chi)=U(\chi)+\delta U
$$
where
\bea\label{U-Xi}
U(\chi)&=&-\mu^4\left[1-\sqrt{1-\Xi^2}-\Xi\arcsin\Xi\right]\,,\\
\label{deltaU}
\delta U&=&\frac14\frac{1}{(4\pi^2)}\left(\frac{\mu^4}{f^2}\cos\frac{\chi}{f}\right)^2\left(\ln(\frac{\mu^4}{f^4}\cos\frac{\chi}{f})-\frac32\right).
\eea
The one loop correction term $\delta U$ may be compared to the classical term $U$: $\delta U/U\sim \mu^4/f^4\sim H/\mpl$. Therefore, one can safely approximate the potential by its classical value $U(\chi)$. To summarize, one may integrate out the axion field during the slow-roll \emph{occurring around the minimum of the potential} and instead work with a gauge theory whose action consists of  $F^2$ Yang-Mills term and $U(\Xi)$. Here some comments are in order:
\begin{itemize}
\item As pointed out,  for positive $\Xi$ (which we assume), \eqref{axion-minimum} has two roots, $\chi/f$ and $\pi-\chi/f$ and we have chosen $\chi$ to correspond to minimum of the potential.
    The one loop effective action computation has of course been carried out  around this minimum. We have hence in \eqref{U-Xi} replaced $\cos\frac{\chi}{f}$ with $-\sqrt{1-\Xi^2}$.
\item We have crucially used the fact that $\chi$ is slowly-rolling and hence in the effective action dropped its kinetic term. In fact, the detailed analysis of the model reveals that $\dot\psi/H\psi\sim \epsilon^2$ and $\dot\chi/H\chi\sim\epsilon$, where $\epsilon=-\dot H/H^2$ is the slow-roll parameter and is of order $10^{-2}$ for this model \cite{gauge-flation-short}. We also note that the contribution of the Yang-Mills term to the energy density of the system is of order $\mu^4\epsilon$, i.e. inflation is mainly driven by the $U(\Xi)$ and hence $3H^2\mpl^2\simeq U(\Xi)$.\item
    Although both $\psi$ and $\chi$ fields have slow-roll, their rolling is ``adiabatic'' in the sense that during slow-roll inflation $\chi$ and $\psi$ fields vary such that the minimizing condition \eqref{axion-minimum} holds.
\end{itemize}

If we are in $\Xi\ll 1$, or equivalently  $\pi f-\chi\ll 1$ regime, one may expand the effective potential, and to second order in $\Xi$ we obtain
\be
U(\Xi)=\frac{\mu^4}{2}\Xi^2+\mathcal{O}(\Xi^4)=\frac{\lambda^2}{128\mu^4} (F\tilde F)^2+\mathcal{O}(\Xi^4)\,.
\ee
Comparing this with the $F^4$ term in gauge-flation model \cite{gauge-flation-short,gauge-flation-long}, we can read its coefficient $\kappa$
\be\label{kappa-mu}
\kappa=\frac{3\lambda^2}{\mu^4}\,.
\ee

In other words, in the $\Xi\ll 1$ regime the chromo-natural inflation model simply reduces to the gauge-flation model.\footnote{For the $\Xi\ll 1$ one may directly expand the axion potential around $\pi f$: $\chi=\pi f+\phi$. Then the $\phi$ field is a ``massive axion'' field of mass $M=\frac{\mu^2}{f}$. As discussed in \cite{gauge-flation-short,gauge-flation-long}, assuming that $H\lesssim M \ll \mpl$ and that $M\simeq \Lambda$, where $\Lambda$ is the UV cutoff of the theory, one may integrate out the massive axion to obtain gauge-flation.}
As discussed in \cite{gauge-flation-short, gauge-flation-long}, the successful inflation model happens for the following values of the parameters and initial conditions of the axion action \eqref{chromo-narutal-action}:
\be\label{gauge-flation-parameter-range}
\begin{split}
g&=(0.4-2.5)\times 10^{-3}\,,\quad \psi=(3.5-8)\times 10^{-2}\,,\\
H&=3.5\times10^{-5}\,,\quad \epsilon=(0.9 - 1.2)\times 10^{-2}\,,\\
\eta&=\psi^2\,,\quad {\kappa} \simeq \frac{2}{g^2\psi^6}\,,\quad \lambda\sin\frac{\chi}{f}=\sqrt{2\kappa} H\,,
\end{split}
\ee
where $\epsilon$ and $\eta$ are the slow-roll parameters.
In this case \eqref{axion-minimum} reduces to $\lambda\sin\chi/f\simeq 10^2$. Since we are demanding that $\pi-\chi/f\ll 1$, $\lambda$ should be large. We may choose $\sin\chi/f=0.05$ as a typical benchmark, leading to $\lambda=2\times 10^{4}$. For these values $\mu=4\times 10^{-2}$. $f$ may consistently be chosen to be equal to $\mu$ and hence $\chi\simeq 2\times 10^{-3}$.

For the gauge-flation model, as discussed, during inflation the dimension 8 operator $(F\tilde F)^2$ dominates over the Yang-Mills term and one may wonder if the one-loop effective action description for gauge-flation discussed above is a valid one. The above analysis clarifies this issue: The $(F\tilde F)^2$ term is coming from integrating out a massive axion and not from gauge interactions and the energy density of the axionic field is of the order the $(F\tilde F)^2$ term which drives inflation. It also makes apparent that, despite being a dimension 8 operator, why one can safely ignore the other gauge invariant dimension 8 operators and possibly higher dimension corrections to the Yang-Mills term.
For example, at dimension 8 level, if we have fermions of mass $m$ in the model  they lead to $Tr F^4$ or $Tr(F^2)^2$ type terms at one loop level. All such terms are, nevertheless,  suppressed by powers of $g^4/m^4$. In the region of parameter space suitable for inflationary dynamics, $g^2/4\pi\sim 10^{-7}$ and the fermion mass $m$ that can contribute to such one-loop effects should not be smaller than  $H$. Such corrections  may hence be safely  ignored.
\footnote{As mentioned,
\eqref{axion-minimum} has a solution which maximizes the potential. One may wonder if it is possible to obtain a successful ``hilltop'' type slow-roll inflation expanding the theory around this maximum and to get slow-roll \emph{only} in the region where $\chi$ is small. Note that although the inflationary trajectory discussed in \cite{chromo-natural} starts when $\chi$ is small, the slow-roll continues for the whole range when $\chi$ becomes close to the minimum of its potential \cite{Mark-Peter}. It is interesting to check whether one can arrange the parameters of the action \eqref{chromo-narutal-action} such that slow-roll ends when $\chi$ moves away from $\chi\ll f$ region. If this is possible, then for the slow-roll inflationary region one may expand the theory around $\chi/f\ll1$ regime and obtain the following effective potential:
$
U\simeq \mu^4\left(2+\frac{\Xi^2}{2}\right)\,,\ \frac{\chi}{f}\simeq \Xi\,.
$}

\textbf{\emph{{Chromo-natural inflationary trajectory, starting around maximum of the axion potential}}}---As has been shown in \cite{Mark-Peter} the action \eqref{chromo-narutal-action} allows for slow-roll inflationary trajectories starting from arbitrary $\chi$ (not necessarily small or close to $\pi f$). This happens with the following set of the parameters \cite{chromo-natural}
\be\label{chromo-natural-parameter-range}
\begin{split}
&\mu^2=10^{-7}\,,\quad f=10^{-2}\,,\quad H\simeq \sqrt{\frac23}\mu^2=8\times 10^{-8}\\
&\lambda=200\,,\quad g=2\times 10^{-6}\,,\quad \epsilon\simeq 1.8\times 10^{-3}\,,\\
&\eta\simeq 1.4\times 10^{-3}\,,\quad \dot\psi\sim -1\times 10^{-6}H,\quad \psi\simeq 3.3\times 10^{-2}\,.
\end{split}
\ee
(For the above values of parameters to get enough number of e-folds, inflation should start when $\chi_0/f< 0.4\pi$ \cite{Mark-Peter}, while it could be chosen arbitrarily in $(0,0.4\pi)$ range.) In the above and also in \eqref{gauge-flation-parameter-range} all dimensionful quantities have been measured in units of $\mpl$.

\textbf{\emph{Comparison between the two cases--}}
We have discussed that the model described by the action \eqref{chromo-narutal-action} indeed allows for some different slow-roll inflationary paths, for different regimes of the parameters of the theory $(\mu,\lambda,f, g)$. Two such cases were given in   \eqref{gauge-flation-parameter-range} and \eqref{chromo-natural-parameter-range}. The gauge-flation case occurs in a ``large axion'' range while the chromo-natural model happens when axion starts with a smaller value, e.g. smaller than $0.4\pi f$ for the parameters in \eqref{chromo-natural-parameter-range}. The chromo-natural trajectory happens for a wider range over possible initial conditions for axion field and is hence more generic in this respect. As discussed, in either case during  slow-roll the extremizing condition \eqref{axion-minimum} holds. Existence of the above two possibilities indicates that one should be able to increase the $\chi_0/f<0.4 \pi$ bound by changing the parameters of the model, essentially filling the gap between the parameters set in \eqref{gauge-flation-parameter-range} and \eqref{chromo-natural-parameter-range}. It is interesting to study this possibility in more detail, some first steps in this directions has been taken in \cite{Mark-Peter}.

For both of the parameter sets in \eqref{gauge-flation-parameter-range} and \eqref{chromo-natural-parameter-range}
$\mu$ and $f$ are of order GUT scale $10^{15}-10^{16}$ GeV. For the gauge-flation case, however, we need a bit lower gauge coupling $g\sim 10^{-3}$, while for the chromo-natural case $g\sim 10^{-6}$. On the other hand, for gauge-flation we need to take a larger value (by one or more orders of magnitude) for the dimensionless axion coupling $\lambda$ compared the one required for chromo-natural inflation.    Moreover, the gauge-flation case has a higher value of Hubble $H\sim 10^{-5}\mpl$ and much bigger tensor-to-scalar ratio $r\simeq 0.1$, while for chromo-natural case $H\sim 10^{-7}\mpl$ and $r\simeq 10^{-6}$ \cite{chromo-natural}. In this respect, gauge-flation range of parameters has a more interesting observational prospects.

Dealing with vector gauge fields at background level one may worry about stability of the inflationary trajectory against anisotropic initial conditions. Such an analysis for gauge-flation was carried out in \cite{Jiro-gauge-flation} and it was shown that the isotropic FLRW inflationary path is indeed an attractor of the model. The main reason behind this behavior is related to the specific form of the $(F\tilde F)^2$ term and that the energy momentum tensor resulting from this term, which is the dominant contribution to the energy momentum tensor driving inflation,  does not contribute to anisotropic stress. Similar reasoning and result also applies to the  path discussed in chromo-natural inflation model  \cite{chromo-natural}.


I would like to thank Peter Adshead and Mark Wyman   for useful correspondences, clarifying comments and sharing the draft of their upcoming article, and Amjad Ashoorioon for comments. I would like to  especially thank Azadeh Maleknejad for
useful discussions and comments.




\end{document}